\documentclass[natbib]{svjour3}                     
\smartqed  
\usepackage{graphicx}
\begin{document}

\title{Multi-wavelength variability
}
\subtitle{Accretion and ejection at the fastest timescales}


\author{    Phil Uttley     \and
     Piergiorgio Casella
}


\institute{P. Uttley \at
              Astronomical Institute ``Anton Pannekoek'', University of Amsterdam, Science Park 904, 1098 XH Amsterdam, the Netherlands \\
              \email{p.uttley@uva.nl}
           \and
           P. Casella \at
              INAF, Osservatorio Astronomico di Roma, Via Frascati 33, I-00040 Monteporzio Catone, Italy \\
             \email{piergiorgio.casella@oa-roma.inaf.it}
 }

\date{Received: date / Accepted: date}

\maketitle

\begin{abstract}
Multiwavelength variability data, combined with spectral-timing analysis techniques, provides information about the causal relationship between different physical components in accreting black holes.  Using fast-timing data and long-term monitoring, we can probe the behaviour of the same components across the black hole  mass scale.  In this chapter we review the observational status of multiwavelength variability in accreting black holes, from black hole X-ray binaries to AGN, and consider the implications for models of accretion and ejection, primarily considering the evidence for accretion disc and jet variability in these systems.  We end with a consideration of future prospects in this quickly-developing field.

\keywords{Multi-wavelength \and Variability \and Black holes}
 \PACS{04.70.Bw \and 95.75.Wx \and 97.60.Lf \and 97.80.Jp \and 98.62.Js \and 98.62.Mw}
\end{abstract}

\section{Introduction}
\label{intro}

A wealth of observational results reveal how time-averaged spectra from variable objects, effectively `wash away' a large amount of relevant information contained in the variability. The emission from the different components of the accretion flow/outflow is now known to change rapidly over a wide range of timescales, as short as minutes in the lowest-mass AGN and milliseconds in the black-hole X-ray binaries (BHXRBs). Resolving such variability down to the shortest possible time-scales is crucial in order to fully disentangle the physical processes causing the emission.

The different emitting components are in fact necessarily inter-connected via the inflowing/outflowing matter itself, at least some fraction of which is likely to be transiting from one component to the other. As different physical components emit at different wavelengths, multi-wavelength studies can offer the unique opportunity to observe the mass flow through the different components of the accretion inflow (disc, hot flow/corona) and outflow (jet, wind). 

This is especially true for those spectral components whose energy budget is somewhat connected to their internal variability, as is the case for the jet, and those whose origin is still unclear, as is the case of the X-ray power-law component. Different works describe the latter in terms of inverse-Compton emission from hot inflowing electrons (Poutanen, this volume) or from the base of an outflowing, collimated jet (Pe'er, this volume). Despite the many physical differences between these models, the degeneracy between them is still large and currently unresolved. Multi-wavelength variability studies may prove to be key here, as the component(s) emitting the photons in X-rays (the energy range where most of the variability is observed and traditionally studied, see Sect.~\ref{sec:pow:most} and Belloni and Stella, this volume), could very well (and probably does) emit at longer wavelengths as well. Thus, simultaneous observations over the full energy range of the emission, i.e. both at optical and X-ray wavelengths, might allow us to solve this conundrum.

Furthermore, we can use correlated multi-wavelength variability to search for lags between bands and also study the time-lags within a waveband (e.g. between soft and hard X-rays) to study the {\it causal} relationship between the different components in the spectrum.  Huge observational progress has been made in this area in recent years, both in the measurement of correlations between X-ray and optical/IR emission, in both AGN and X-ray binaries, as well as the detailed study of time-lags within the X-ray band, which has also led to the discovery of X-ray reverberation (see also Reynolds, this volume).  We will review this observational progress in this chapter, and also give a flavour of the models to explain the observed behaviour, which are still at an early stage but already show significant promise as a means to understand the physical structure of the emitting regions close to black holes.  We also point the reader to the other relevant chapters in this volume (e.g. Pe'er, Poutanen) for a more detailed look at some of the models and associated issues.

\subsection{Stellar-mass vs. supermassive BHs: complementary information}
\label{intro:mass}

The study of the physics of the accretion onto BHs cannot overlook the issue of the similarities and differences between accretion on to stellar-mass BHs in XRBs and supermassive BHs in AGN. On one hand, it is reasonable to assume that the physics at play is substantially the same. On the other hand, some differences are expected, associated with the different energy densities at work. In fact, the very existence of BHs over such a broad range of masses can be seen as an outstanding opportunity to tackle these studies from two different, complementary perspectives. In AGN, since their black holes are $\sim 10^{5-8}$ times more massive than those in BHXRBs, the intrinsic luminosities (at the same fraction of the Eddington luminosity) are very high, and the characteristic timescales are very long compared to those in BHXRBs, from minutes for the shortest dynamical times, to years or longer for viscous time-scales.  Thus, at least for the nearest AGN, the number of photons per characteristic timescale is typically several orders of magnitude higher than in stellar-mass BHs, allowing us a high-precision look at the detailed physics of individual variations on the fastest variability time-scales. 

On the other hand, stellar-mass BHs have much shorter characteristic timescales (viscous, dynamical or thermal), so that data can be collected over large numbers of cycles to give an excellent picture of their average variability properties, which is extremely useful when the variability processes are themselves stochastic, as they are for accreting black holes.  We can thus probe - with high precision in just a few hours - the signals associated with viscous evolution of the accretion flow which would be observed over only a few cycles in 10 years of AGN monitoring.   Crucially also, in BHXRBs we are able to observe the long-term evolution of the accretion flow through different spectral states, which may well be realised in AGN on much longer (and cosmologically relevant) time-scales (see also Koerding; Fender \& Gallo, this volume).  Using multiwavelength techniques on observations of the same BHXRB in different states can thus allow us to unlock the changes in the inner structure of accreting black holes and connect them to long-term behaviour such as jet and wind formation: again, a topic of key relevance for the study of supermassive black holes and their feedback into the environment.

\subsection{Multi-wavelength variability tools and techniques}
\label{intro:tools}

In this Chapter, we will review the state of the art of multi-wavelength variability of accreting BHs. In particular, we will focus on the most recent results obtained with two relatively new methods/strategies as well as a tried-and-trusted (but previously difficult to do) old strategy. On the one hand, the use of novel timing tools, like the covariance spectrum \citep{Wilkinson2009} and the detailed study of lag vs. energy (e.g. \citealt{Uttley2011,Kara2013}), together with the large-area and soft-response of satellites like {\it XMM-Newton}, are allowing us for the first time to identify and study the disc variability driving the variability of the hard X-ray component as well as measuring the signature of reverberation of the variable power-law off of the disc, thus broadening our interpretative approach.  These methods simply extend the traditional Fourier cross-spectral techniques to measure correlations and lags (see \citealt{Nowak1999} for a useful review) to look at the causal relationship between much finer energy bins relative to a broad reference band, chosen to optimise signal-to-noise.  

At the same time, high time resolution observations have started - in the early 2000s - to become available at wavelengths different than the "traditional" X-rays, after the first isolated results obtained with proportional counters a few decades ago \citep{Motch1983}. Dedicated fast optical photometers with good quantum efficiency \citep[e.g.][]{Kanbach2003,Dhillon2007} as well as fast modes implemented in existing near-infrared detectors, allow us to tackle old questions from a new perspective. The fastest timescales of the accretion onto stellar-mass BHs are now accessible for studying in the synchrotron emission from both the very same electron population responsible for the X-ray Comptonized emission, and - at longer wavelengths -  the electron population outflowing in the relativistic jet.

Finally, the previous decade saw significant improvements in the availability of more traditional long-term monitoring data.  Before the launch of the {\it Rossi X-ray Timing Explorer} ({\it RXTE}), high-quality X-ray monitoring was near-impossible to achieve, since previous observatories were not set up for the flexible scheduling and fast-slewing to targets which allow efficient monitoring.  After the launch of {\it RXTE}, X-ray monitoring to study the detailed long-term variability of AGN (as well as catching X-ray transients in different stages of their outbursts) became much easier.  At the same time, optical monitoring campaigns became easier to obtain, thanks to the work of dedicated groups with access to institutional telescope time, as well as the development of a new generation of smaller - sometimes robotic - telescopes (e.g. the Liverpool Telescope and the SMARTS telescopes), which are ideal for carrying out photometric monitoring if AGN.  Finally, the launch of {\it Swift} satellite has allowed combined X-ray and optical/UV monitoring from space, adding another string to the multiwavelength-monitoring bow.

\section{X-ray power-law variability}
\label{sec:pow}

The X-ray power-law continuum component is the best-studied with spectral-timing techniques, but arguably still the least understood, given the uncertainties in the physical origin of this component.  In this section, we will discuss what is known about the spectral-timing behaviour of the X-ray power-law and then describe the attempts to model and understand this behaviour.  We shall focus throughout this chapter on the broadband noise variability, rather than quasi-periodic behaviour, since more substantial progress has been made in understanding the spectral-timing behaviour of the broadband noise, and clearer connections can be made between BHXRBs and AGN, which only show one clear example of a QPO to date \citep{Gierlinski2008}.  However, we note here that the QPO variability seems to be primarily associated with the power-law component and may be produced by a distinct physical mechanism from that producing the broadband noise, as discussed by Belloni \& Stella (this volume).

\subsection{The most variable spectral component?}
\label{sec:pow:most}

The X-ray band shows some of the most rapid, largest-amplitude variability of all wavebands from accreting objects, including both black hole X-ray binaries and AGN, although the variability time-scales themselves need to be corrected for the black hole mass (see Koerding, this volume).  For many years, it has been realised that the power-law emission dominates this variability.  

In AGN, it is clear that the X-ray power-law is the dominant emission component in the X-ray band, since with the exception of a few low-mass, high accretion rate AGN, the disc should emit primarily in the optical/UV.  Various spectral variability techniques, such as time-resolved spectroscopy \citep{Markowitz2003}, simple flux-flux plots \citep{Taylor2003,Ponti2006} as well as rms-spectra \citep{Vaughan2004,Papadakis2007,Arevalo2008a}, show that flux variability in AGN is dominated by changes in normalisation of the power-law component, sometimes with no corresponding change in power-law spectral shape.  There is some evidence for variable absorption driving some of the observed X-ray variability in AGN, most notably on short, sub-day time-scales in the `eclipses' seen in the now-famous case of NGC~1365 (e.g. \citealt{Risaliti2005}), but also on longer time-scales (weeks to months) in other AGN (e.g. \citealt{Lamer2003,Rivers2011}).  These absorption variations are generally interpreted in terms of the observer's line of sight passing through the so-called Broad Line Region (BLR) which produces the broad optical emission lines seen in type 1 AGN, or the inner edge of the dusty torus just beyond the BLR.  The BLR can act as a variable partial covering screen or more appropriately a `mist', since the BLR clouds may span a range of size scales and may have a smaller physical size than the central X-ray emitting region.  

There are also claims that the larger-amplitude X-ray variability and 'harder-when-fainter' X-ray spectral changes observed in a large number of AGN are driven at least partly by absorption by a combination of variable partial covering and reflection, from material as close as 100 gravitational radii (light-minutes-hours) from the black hole \citep{Legg2012,Tatum2013}. This possibility is arguably most convincing in sources which show the most extreme spectral variability \citep{Miller2009,Lobban2011}, although this is hotly debated even in these cases, with much of the evidence centering on the interpretation of the X-ray time delays seen in AGN (e.g. see \citealt{Miller2010,Zoghbi2011} and Reynolds, this book).  However, given the ubiquity of AGN variability even where there is little spectral evidence for variable absorption, the similarities between AGN and BHXRB variability properties and the correlations between the optical and X-ray continuum variability, which we outline below, it seems likely that the variability in most AGN is primarily intrinsic and not driven by absorption.  We will therefore assume that this is the case in the remainder of this section.

In BHXRBs, the X-ray band contains emission from both the disc blackbody and the power-law, although in the hard states the disc emission can only be studied with CCD detectors that extend spectral coverage to energies $<2$~keV \citep{Miller2006,Wilkinson2009}.  For this reason, the study of spectral-timing in BHXRBs, pioneered by satellites with proportional-counter detectors such {\it Ginga} and later {\it RXTE}, has necessarily been biased towards studying the power-law in the hard states (e.g. \citealt{Nowak1999,Revnivtsev2001}).  In the softer states on the other hand, both the power-law and disc components can be studied and there it is clear that the power-law dominates the variability and disc emission seems to be rather stable.  For example, rms-spectra of the soft state of the high mass BHXRB Cyg~X-1 show clearly that the power-law varies but not the disc emission \citep{Churazov2001}.  This effect could help to explain why the broadband fractional rms of BHXRBs decreases as they get softer (e.g. \citealt{Belloni2005}): the weakly-varying/constant disc component becomes stronger and this dilutes the intrinsic power-law variability amplitude.

\subsection{Time-lags}
\label{sec:pow:lags}

The first Fourier cross-spectral study of the time lags in accreting black holes \citep{Miyamoto1988} showed that the lags in hard state BHXRBs were hard (i.e. variations of hard photons lag those in softer photons) and also dependent on Fourier frequency, i.e. variability time-scale.  Specifically, the phase lag increases weakly with frequency, approximately as $\phi \propto \nu^{0.3}$, or equivalently, {\it time lag} scales as $\tau \propto \nu^{-0.7}$.  Typical hard-state hard lags are around 1~per~cent of the sampled variability time-scale (e.g. between 3--4~keV and 8--14~keV, \citealt{Nowak1999}), so the effect is relatively small, but despite that, the time lags at the lowest well-sampled ($\sim$mHz) frequencies are in excess of thousands of the light-crossing time for one gravitational radius.  The lag between two energies scales roughly linearly with the logarithmic energy separation \citep{Nowak1999}, although the best measurements to date of the lag-energy dependence of the power-law component indicate a more complex shape than a simple log-linear law, notably with  `wiggles' seen around the energy of the iron K$\alpha$ line \citep{Kotov2001}.

The above-noted properties are well-known, but another less-often noted but important feature of the hard state lags is that, although they follow an approximate power-law dependence on Fourier-frequency when measured over several decades in frequency, the best measurements indicate that they show a step-like structure, with the time-lags appearing to be relatively constant over broad ranges in frequency before `stepping' to a lower lag value.  As noted by \citet{Nowak2000,Kotov2001}, the steps appear to correspond to the frequencies where the dominant component in the power spectrum switches from one broad Lorentzian to another.  In other words, each variability component in the power-spectrum appears to have its own, roughly constant, time-lag, with the time-lag increasing as the time-scale of the component increases.
\begin{figure*}
  \includegraphics[width=0.53\textwidth]{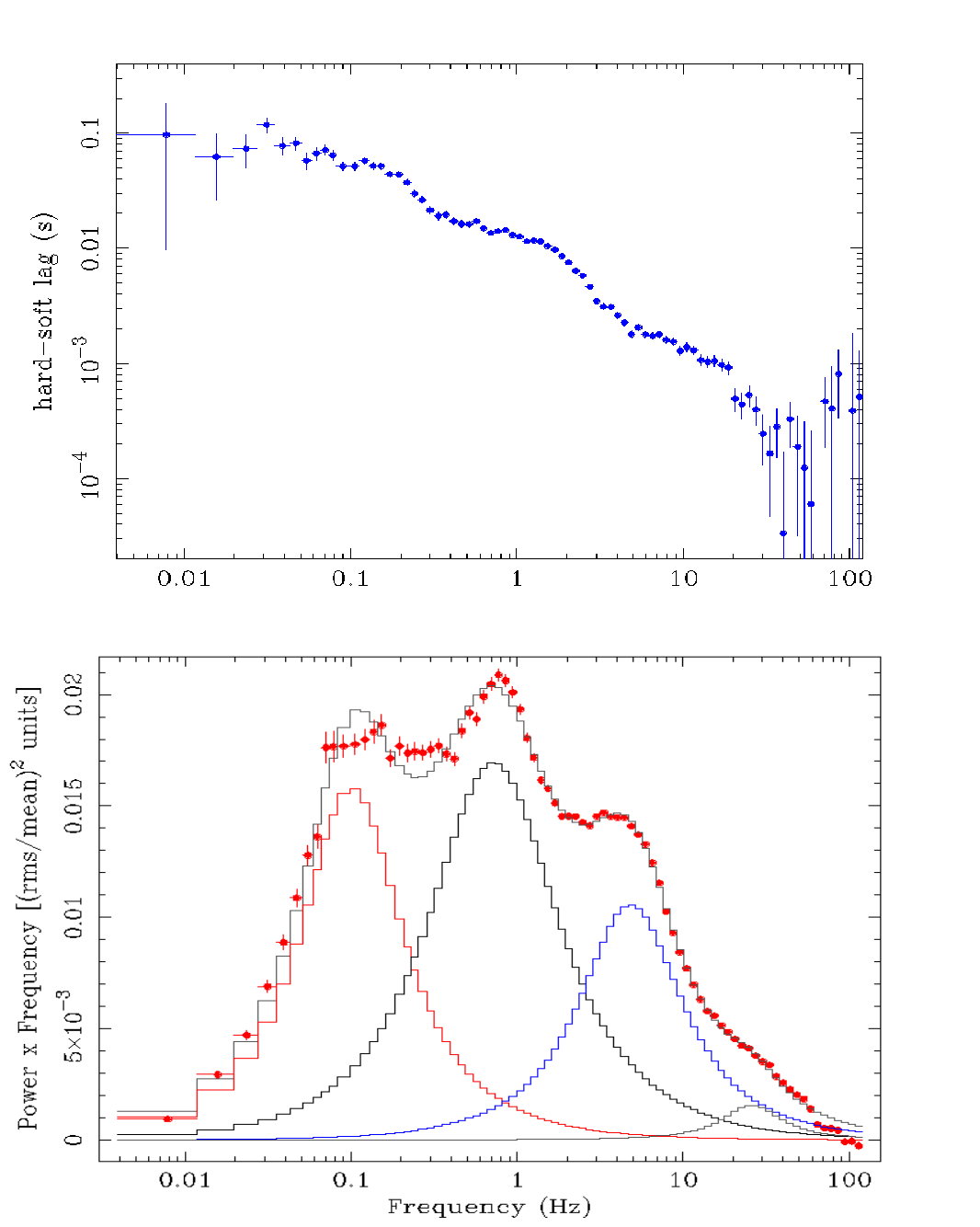}
  \includegraphics[width=0.45\textwidth]{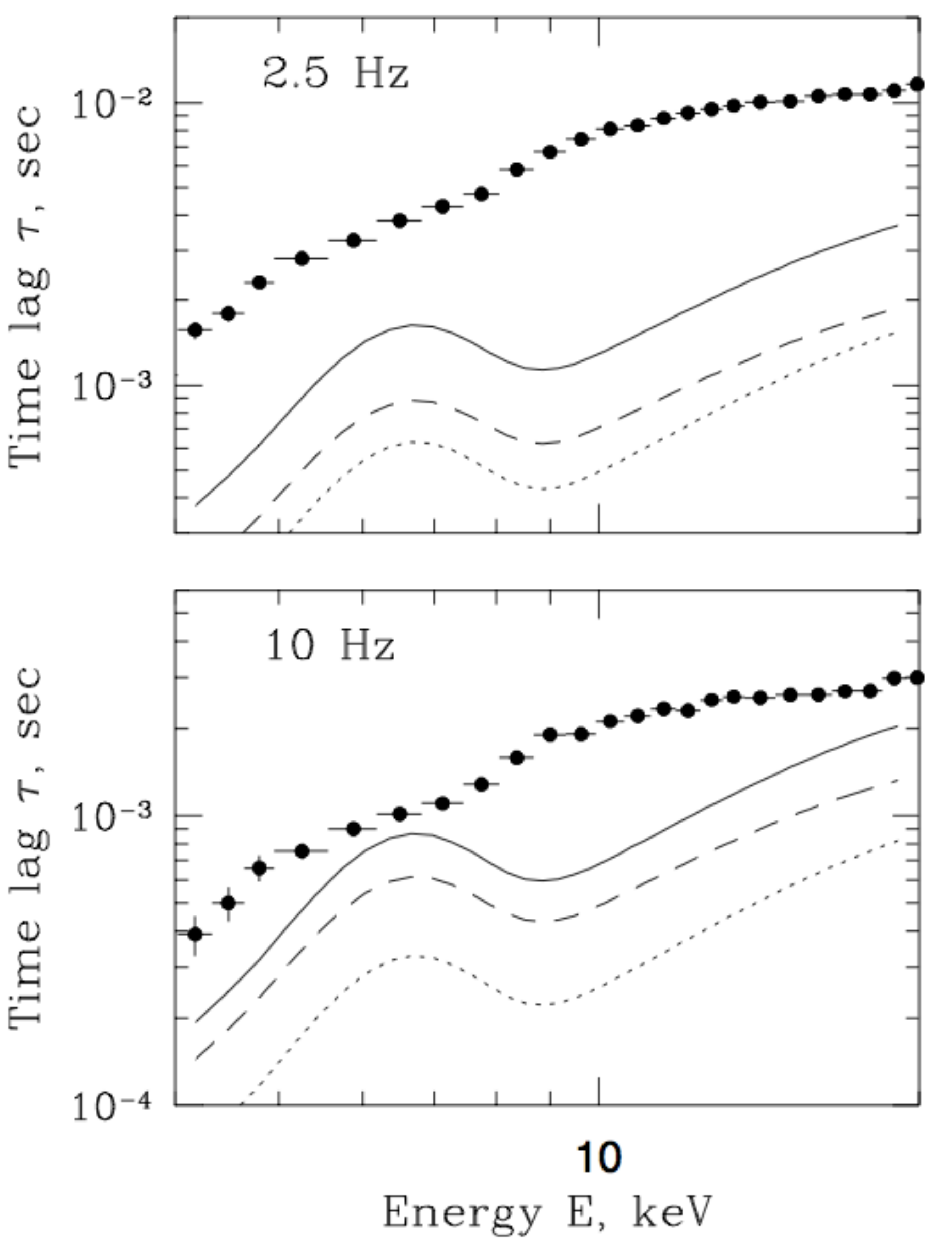}
\caption{Left: The top panel shows the 8--13~keV vs 2--4~keV time-lag dependence on frequency for Cyg~X-1 in the hard state (observation from December 1996).  Although the general trend follows $\tau \propto \nu^{-0.7}$, the `stepped' nature of the lags is very clear.  The steps appear to correspond to the transitions between the broad Lorentzian components which approximately fit the hard state power spectrum, shown below.  Right: Lag (with respect to variations in the 2.8--3.7 keV band) versus energy, measured for variations at 10~Hz and 2.5~Hz for the hard state of Cyg~X-1.  The figure, taken from \citet{Kotov2001}, also shows the expectations of simple reflection models for the lags.  Although the lags do not look like they are caused by simple reflection, they also clearly do not follow the simple log-linear law expected from Compton upscattering, and show interesting `wiggles' around the Fe~K features.}
\label{fig:xrbrxtelags}
\end{figure*}

The hard-state lags are also seen to evolve with the changes in the power spectrum which correlate with X-ray spectral changes.  The most systematic work so far has been done for Cyg X-1, with \citet{Pottschmidt2000,Pottschmidt2003} showing that the lag in a fixed frequency band increases as the spectral index steepens (and the frequencies of the broad Lorentzian power spectral components correspondingly increase), with large lags seen during the intermediate states and so-called `failed state transitions'.  Interestingly, the soft state of Cyg X-1, where variability is dominated by the power-law spectral component, shows smaller lags, with an amplitude and frequency-dependence similar to the hard state (although without the clear `steps').  The lags in the soft states of transient low mass BHXRB sources are not well-studied to date, due to their much lower variability amplitudes than in Cyg~X-1 (fractional rms $\sim 1$\% or less versus $>10$\% in Cyg~X-1), and their power-law spectral components are correspondingly much weaker relative to the disc emission than in Cyg~X-1.

Studying the lags in AGN to equivalent (mass-scaled) frequencies as sampled in BHXRBs is not yet possible, since the intermittent sampling that was used to monitor long-time-scale AGN variability is not sufficient to measure lags which are only a small fraction of the observed variability time-scales.  However, days-long, intense (quasi)continuous observations can probe the lags on shorter time-scales (equivalent to $\sim 1$ second or less in BHXRBs).  Fourier-frequency-dependent time-lags in AGN X-ray variability were first observed in data from {\it RXTE} observations of the Seyfert galaxy NGC~7469 
and later in {\it XMM-Newton} and {\it ASCA} data for several other AGN (e.g. \citealt{Vaughan2003,McHardy2004,Arevalo2006b}).  

These results indicated a similar picture to that seen in hard state BHXRBs (and Cyg X-1 in the soft state), with hard time-lags decreasing inversely with variability time-scale (consistent with $\tau \propto \nu^{-0.7}$ but with large uncertainty) and following an approximately log-linear scaling with energy.  Ark~564, the only AGN known which shows a doubly-broken, or more likely multi-Lorentzian, power-spectrum\footnote{Other AGNs show power spectra consistent with singly-broken or bending power-laws, similar to those seen in BHXRB soft states, e.g. \citet{McHardy2004}.}, added to the similarity with BHXRBs when combined short and mid-time-scale data revealed a sharp drop to a possible negative (i.e soft) lag at a frequency consistent with the change-over in power-spectral components \citep{McHardy2007}.  Subsequently, following the first highly significant detection in 1H0707-495 \citep{Fabian2009}, soft high-frequency lags have been confirmed in {\it XMM-Newton} data from many other AGN (e.g. \citealt{DeMarco2013}), and these are now considered to be associated with reverberation from relatively close to the black hole, which we will not discuss further here (but see Reynolds, this volume).  The origin of the hard lags observed in AGN at lower Fourier frequencies is still debated, however.

\subsection{Models for longer time-scale lags: light-travel times, or something slower?}
\label{sec:pow:models}

When measuring time-lags it is implicit that there is a correlation (or anti-correlation for large phase lags) between the different energy bands, i.e. the variability is {\it spectrally coherent} to some degree.  A correlation implies either a direct causal connection between the  mechanisms producing the variations in photons in each band, or a causal link of both bands to some underlying variable process.  The lag contains causal constraints on these physical situations and the amplitude of the lag is strongly determined by the speed with which signals produce variations in each energy band.  The fastest speed which can be considered is light-speed and many models for lags invoke light-travel times as the primary origin of the lag.  

Early proposals for the physical origin of lags in BHXRBs focussed on Compton upscattering, where lags are imparted by the travel times of photons in the hot upscattering plasma which produces the power-law emission \citep{Payne1980,Kazanas1999}.  Higher energy photons have undergone more scatterings and hence spend more time in the scattering medium, lagging behind the lower-energy photons which escape sooner.  Since the logarithmic change in energy scales with the number of scatterings (as does the time spent in the upscattering region), this model predicts a log-linear energy dependence in the lags, approximately what is observed.  However, the size of the spherical upscattering region would need to be extremely large ($\sim$a light-second across) to explain the large lags seen at low-frequencies, so these models were subsequently discounted on energetics grounds \citep{Nowak1999}.  Such an extended hot ($\sim 100$~keV) plasma could not be maintained at such scales.

A more plausible Compton upscattering model for lags was proposed by \citet{Reig2003} and subsequently developed by \citet{Giannios2004,Kylafis2008}. The basic mechanism (light-travel times associated with multiple scatterings) is the same as for spherical scattering regions, but the energetics problem is avoided by positing that the upscattering takes place in the jet, which is inferred to extend to large scales, based on the flat-spectrum radio emission seen in the hard state.  Seed photons can be trapped in the jet by being preferentially upscattered in the forward direction, so that they can upscatter several times before leaving the jet, thus imparting the energy-dependent lags. One difficulty with this model is that one should expect a significant fraction of the X-ray continuum emission to be produced hundreds of gravitational radii from the disc, so that the illumination pattern should be relatively flat compared to the disc-like emissivity or steeper profiles inferred from spectral fitting of the Fe~K$\alpha$ line (e.g. see Reynolds, this volume).

The fundamental difficulty with models which invoke light-travel time to explain the low-frequency hard lags in BHXRBs is that the inferred sizes of the emitting regions are too large to be physically plausible and/or consistent with other aspects of the data.  To address this problem, other types of slower signal were considered, most likely associated with the accretion flow itself.  One possibility is the cylindrical propagation of waves or other fluctuations through a corona or hot flow with a temperature that increases with decreasing radius, to produce hard lags \citep{Nowak1999}. 
 A likely scenario seems to be that the lags are associated with the propagation of mass accretion fluctuations \citep{Kotov2001,Arevalo2006b}, since the same scenario can explain many other timing properties, including the linear rms-flux relation observed in accreting sources \citep{Uttley2001,Uttley2005}.  Thus the large lags could be associated with the relatively slow, viscous propagation speeds of signals, meaning that the signals themselves may originate from relatively small regions, within 100~$R_{\rm G}$ of the black hole.

Although the propagating accretion fluctuation model can qualitatively explain the large lags observed at low frequencies in BHXRBs, the situation is less clear for AGN.  Arguably the AGN lags, being of similar relative magnitude and time-scale dependence as the BHXRBs, could plausibly share the same origin as the BHXRB lags.  However, even if that is true, AGN lag measurements do not yet extend to long enough time-scales to effectively rule out light-travel time origins for the lags.  Thus the origin of the hard lags in AGN is currently a matter of debate, with light-travel time associated with reflection being the major alternative to a slower signal propagation effect (e.g. see \citealt{Zoghbi2011,Legg2012}).  It is worth noting that evidence is currently emerging that the reflection signatures expected in the lag vs. energy spectrum are only seen in the high-frequency lags, not at lower frequencies \citep{Kara2013}, and furthermore, there is evidence that even AGN with apparently weak reflection also show hard lags \citep{Walton2013}.  Thus, it seems that the low-frequency hard lags do not show any evidence that they are associated with reflection.  The same is certainly true for BHXRBs, where the evidence against reflection as the origin of hard low-frequency lags is even more compelling \citep{Cassatella2012a}.

\section{Disc blackbody variability}
\label{sec:disc}

\subsection{AGN}
\label{sec:disc:agn}

The optical continuum in AGN has been known to be significantly variable almost since the discovery of these enigmatic objects.  This variability can be clearly associated with the AGN accretion disc, which is the most likely physical origin of the `big blue bump' component responsible for the optical continuum \citep{Koratkar1999}. Studies of the optical variability in different continuum filters showed that variations from the $R$ band through to the UV are very well correlated, with only short {\it red} lags observed, so that longer wavelengths lag shorter wavelengths by up to a day (e.g. \citealt{Wanders1997,Sergeev2005}) .  This behaviour posed a puzzle, because the emission from different wavebands should originate primarily at different radii in a multi-temperature accretion disc.  For intrinsic disc fluctuations to propagate between these regions (either on a viscous or thermal time-scale) should take much longer than the observed sub-day lag times.  Furthermore, one might expect that fluctuations should propagate from the outside in, with accretion variations propagating inwards with the accretion flow, from cooler to hotter parts of the disc. This effect should produce blue lags rather than the observed red lags.  

The likely solution to the puzzle of short optical lags in AGN is to suppose that the optical variability is driven by heating by the central, variable X-ray source \citep{Krolik1991}.  The lags could then be simply interpreted as being due to the differential light-travel time between the emitting regions, with redder bands being emitted in the lower-temperature parts of the disc further away from the central X-ray source, hence short, red lags are observed.  The detailed lag versus wavelength dependence (both amplitude and functional form) is consistent with this picture, assuming a standard disc temperature profile \citep{Cackett2007}.

\begin{figure*}
  \includegraphics[width=0.5\textwidth]{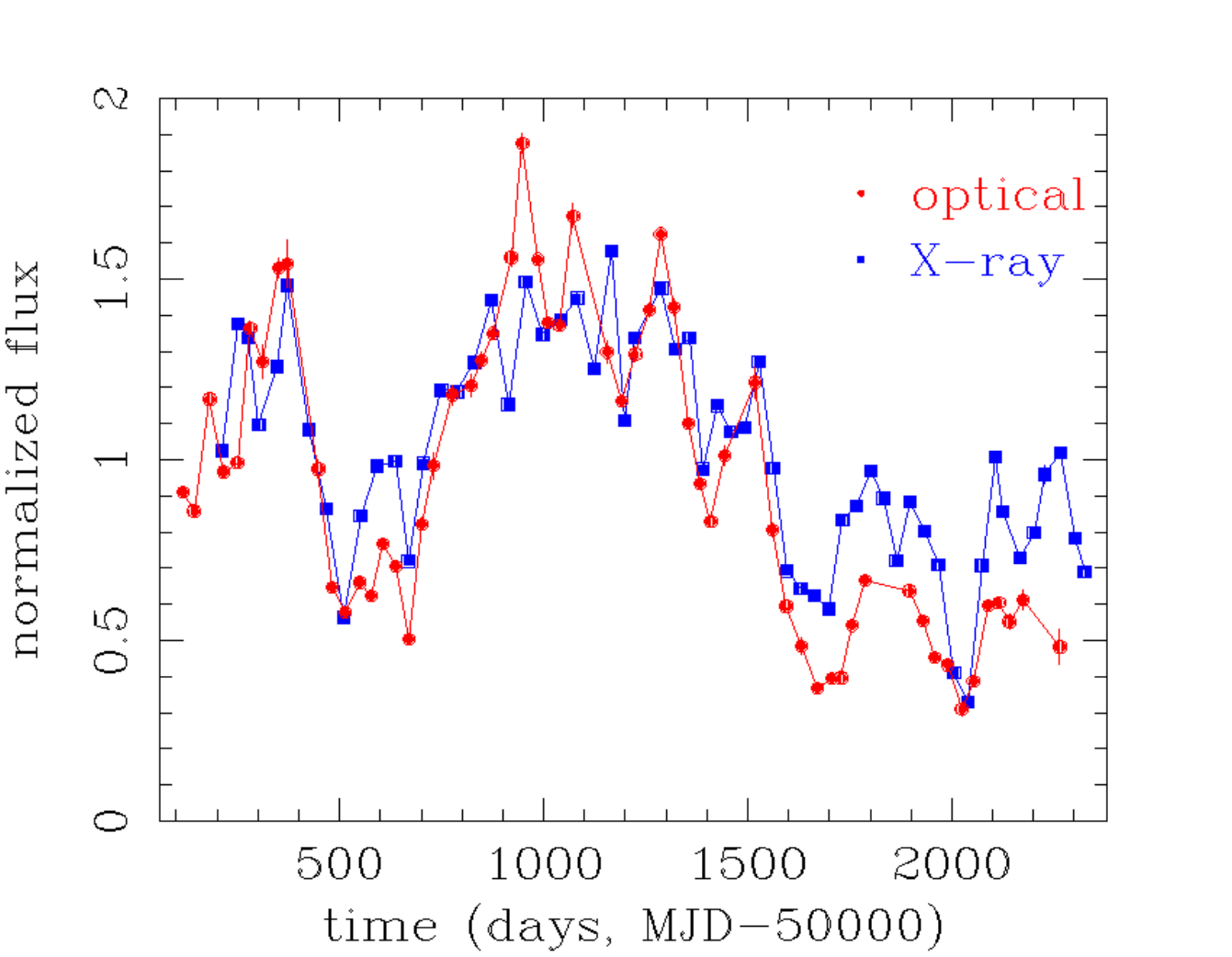}
  \includegraphics[width=0.45\textwidth]{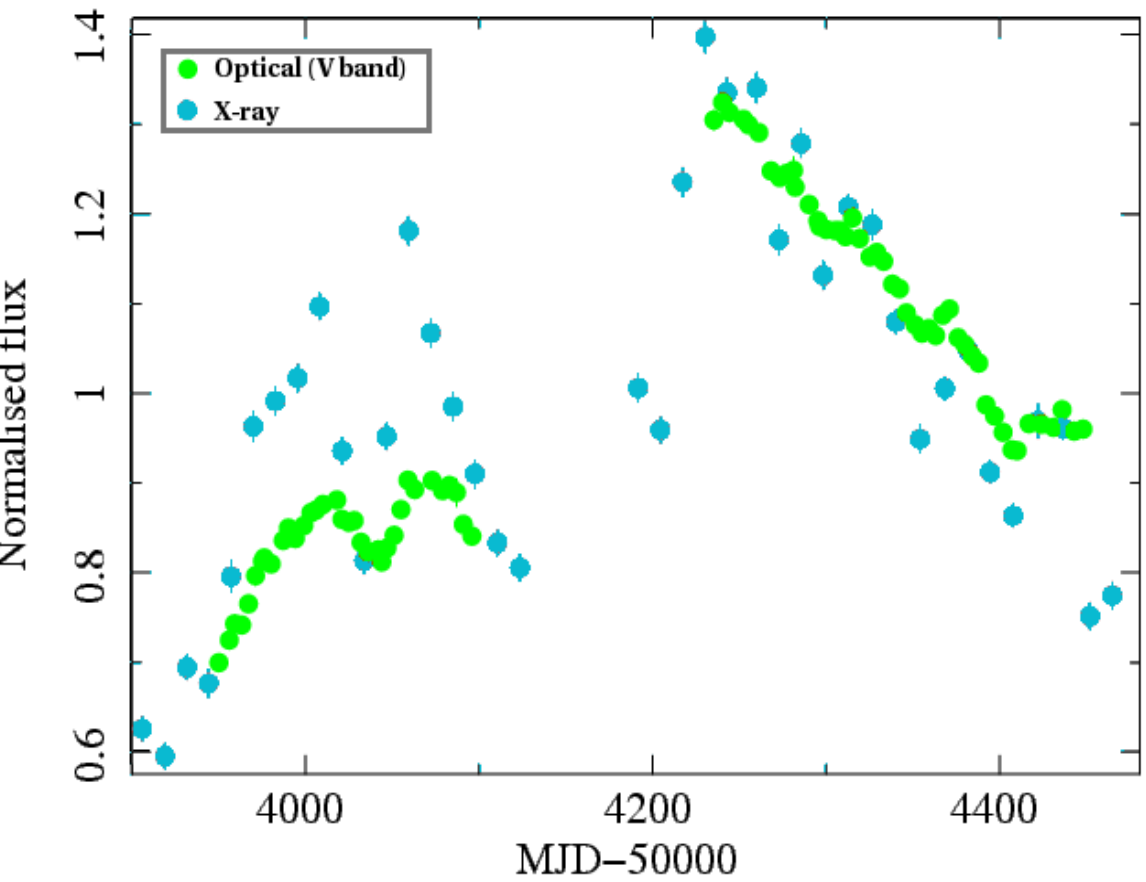}
\caption{Left: Correlated optical continuum (5100 \AA) and 2--10~keV X-ray light curves of the Seyfert galaxy NGC~5548 (adapted from \citealt{Uttley2003}).  The optical light curve has been corrected for the constant host galaxy optical light.  The larger optical variability amplitude implies that long-term optical variations are due to intrinsic disc variability and not driven by X-ray heating of the disc.  This result is strengthened when considering the luminosities in the optical/UV and X-ray components.  Right: Correlated optical and X-ray variability in the quasar MR~2251-178 (adapted from \citealt{Arevalo2008b}).  The large long-term optical variation also implies intrinsic disc variability.  However, the picture is complicated by the small but still correlated optical variations on short time-scales.  These short-term correlated variations with short lags ($<10$~days) can be explained if X-ray heating drives the short-term optical variability from the disc.}
\label{fig:agnxrayoptlcs}
\end{figure*}

The success of the X-ray heating model for explaining AGN optical lags led to an assumption that all AGN optical variability could be explained in this way.  However, a further breakthrough was made once the long-term {\it RXTE} X-ray monitoring programmes used to study Seyfert galaxy power spectra were combined with simultaneous optical monitoring. After an early mixed picture for several AGN \citep{Nandra2000,Peterson2000,Maoz2002}, it became clear that most Seyfert galaxies show a significant correlation between X-ray and optical variations (see Fig.~\ref{fig:agnxrayoptlcs}, left panel).  On short time-scales of days--weeks, the optical variations have significantly smaller amplitude than the corresponding X-ray variations.  However, on longer time-scales of months--years, larger amplitude fractional rms variations are seen in the optical compared to X-rays \citep{Uttley2003,Arevalo2008a,Arevalo2009,Breedt2009,Breedt2010}.  This behaviour is difficult to explain if X-ray heating dominates the variability, since some component of the optical light should be associated with intrinsic disc heating, which if it were constant would dilute the fractional amplitude of variations to a lower value than seen in the driving X-ray emission. The natural interpretation is that a significant component of {\it long term} optical variability is intrinsic to the disc.  The fact that this component is also correlated with long-term X-ray variations further suggests that instabilities in the accretion disc may ultimately drive the X-ray variability on those time-scales.  Perhaps the strongest argument that large-amplitude AGN optical variability is primarily due to intrinsic disc variability, made by \citet{Gaskell2008}, is the fact that the SEDs of many Seyfert galaxies are dominated by the big blue bump, which can be an order of magnitude more luminous than the total luminosity of the X-ray power-law (assuming a standard cut-off energy of $\sim100$~keV).  Combined with the large amplitude of optical variability seen on long time-scales, it seems very unlikely on simple energetics grounds that X-ray heating can be responsible for most of the optical variability\footnote{Note that significant X-ray beaming towards the disc can be ruled out by the observed X-ray reflection component which is not stronger than expected, in most cases.}.

These problems notwithstanding, we still need to explain the short optical lags, and these are best considered by noting that they occur in response to {\it short-time-scale (days--weeks) X-ray variability}.  The response of the disc to rapid X-ray variability dominates the location of the central peak of the cross-correlation function (CCF) which is used to define the lags, rather than the overall CCF shape which is also affected by long time-scale variability.  Putting things together, AGN optical variability may be best explained by a composite picture, where long-time-scale, large amplitude variations are produced by intrinsic variations in the accretion disc (perhaps in mass accretion rate), while short-time-scale smaller amplitude variations (which dominate the optical lag measurements) are driven by X-ray heating (see Fig.~\ref{fig:agnxrayoptlcs}, right panel).  The fact that the long-term X-ray variations are also correlated with long-term optical variations could be explained if the X-ray power-law is also driven by the same mass-accretion variations that drive the optical variability.  If much of the optical variability is driven by mass-accretion variations propagating through the disc, why do we not see the long, blue lags that we expect from such a model?  \citet{Arevalo2008b} address this question by showing how the X-ray heating effect pushes the peak of the CCF to short red lags, while the broad base of the CCF is pushed towards long blue lags by the long-term intrinsic disc variability.  Hence the CCF is asymmetric but still peaks sharply and close to zero, as observed.  The data obtained so far are not yet capable of detecting the asymmetry on long time-scales however.

\subsection{BHXRBs}
\label{sec:disc:bhxb}

As noted above, in the softer states which show SEDs increasingly dominated by disc emission, variability is weak and when significant variability is seen it seems to be dominated by the power-law emission.  These results led to the scenario proposed by \citet{Churazov2001} where the standard, optically thick disc is intrinsically stable and variability seen in the power-law component (including in the hard states) is produced by an unstable, hot, optically thin flow which occurs just inside the disc truncation radius.  The outer and inner radii of this hot flow were suggested as the origins of the low and high-frequency breaks (or equivalently, broad Lorentzians) seen in hard state power spectra.

\begin{figure*}
  \includegraphics[width=0.9\textwidth]{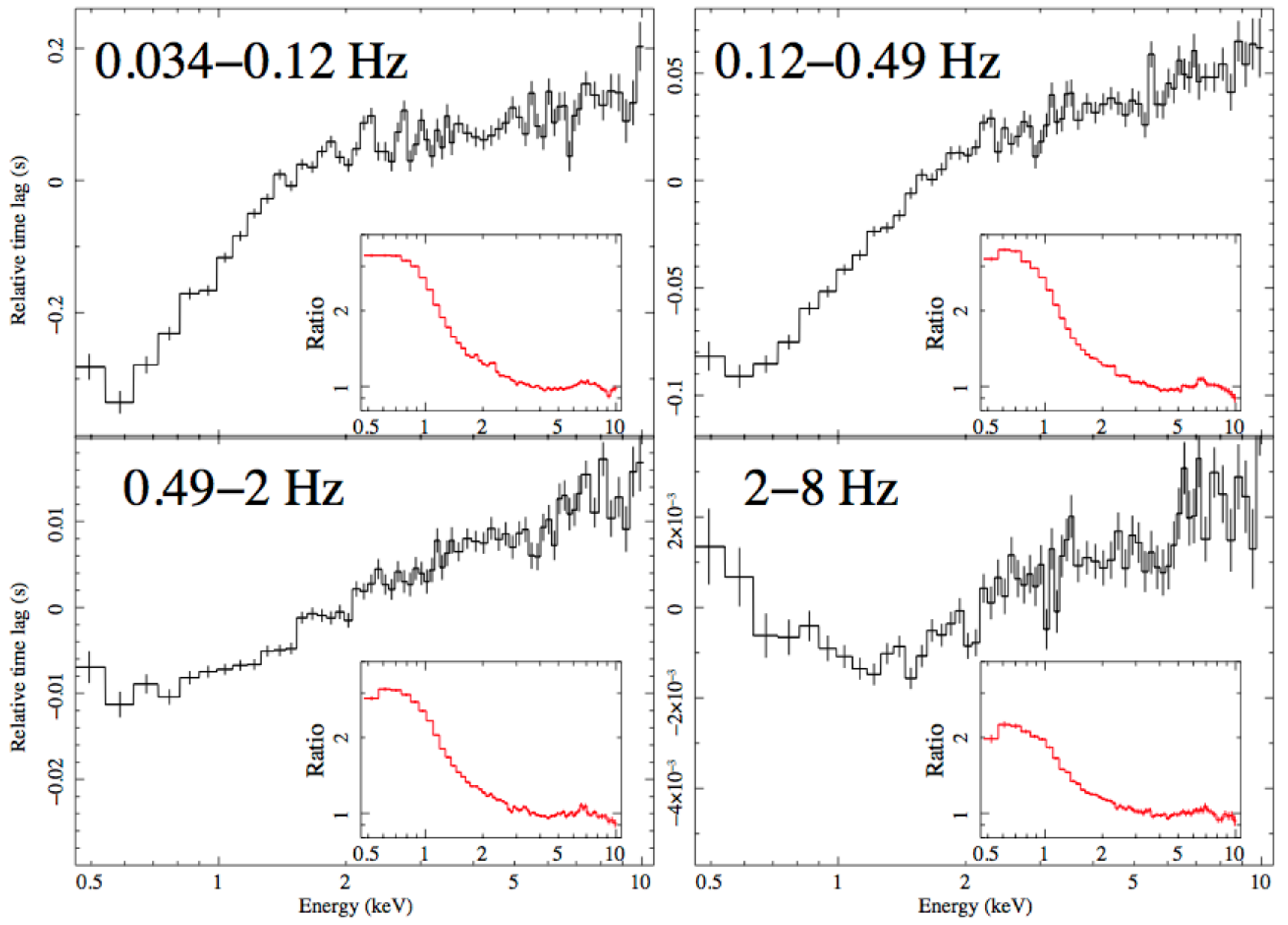}
\caption{Comparison of lag vs. energy for four different Fourier frequency ranges of the X-ray variations of the hard state BHXRB GX~339-4.  The insets show the covariance spectra plotted as a ratio to a Galactic-absorbed power-law, clearly showing as a soft excess the variable disc emission first revealed by \citet{Wilkinson2009}.  The disc photons clearly precede the power-law photons in the lag-energy spectra, by up to tenths of a second at the lowest frequencies.  As the disc variability becomes weaker above 1~Hz, the disc photon lead changes to a short lag, which can be easily explained by the increasing dominance of X-ray heating and associated thermal reverberation lags on these short time-scales.  Figure taken from \citet{Uttley2011}.}
\label{fig:gx339lagvsencomp}
\end{figure*}

One outstanding - and in hindsight, crucial - issue with this picture was that it was based on timing results from proportional counter instruments, notably on {\it RXTE}, which are not sensitive to photons below 2~keV.  Therefore, the actual variability of the cool ($kT<0.5$~keV) discs in the hard state could not be studied.  This problem was solved relatively recently, with the first X-ray timing studies of the hard state which extend to soft X-rays, pioneered with the EPIC-pn instrument on board {\it XMM-Newton}.  These studies made use of advanced spectral-timing techniques, namely the covariance and lag-energy spectra, to show that, not only is the disc substantially variable in the hard state (in fact it can be more variable than the power-law) on time-scales of a second or longer \citep{Wilkinson2009}, but the disc variations {\it precede} power-law variations on these time-scales \citep{Uttley2011}. Variations of the power-law lag the correlated variations of the disc by a tenth of a second or even longer (see Fig.~\ref{fig:gx339lagvsencomp}).  

The hard state disc variability observed by {\it XMM-Newton} has a number of important implications for the origin of the variability as well as the interpretation of spectral-timing data.  Firstly, the obvious implication is that in the hard state, variability (at least on time-scales of seconds or longer) is driven by intrinsic accretion variations in the disc, just as long-term variability appears to be driven by intrinsic disc variations in AGN.  Secondly, the observed lag of the power-law with respect to the disc sets strong and probably fatal constraints on models where the power-law lags are due to Compton scattering, e.g. in the jet.  This is because the disc is almost certainly the source of seed photons, and in these models the large power-law to disc lags would then imply a distance of the scattering region from the disc of tenths of a light-second, i.e. thousands of gravitational radii.  If the X-ray power-law is located such a large distance from the disc, it would surely produce very different reflection spectra than observed.  Furthermore, even if the power-law lags are not due to light-travel times in an upscattering region, the upscattering region cannot itself subtend much of a solid angle as seen from the disc (e.g. the disc cannot be embedded in a corona), or the large lags of the power-law relative to the disc would be washed out since disc photons would be rapidly upscattered to power-law photons.  The implication is that any upscattering region that produces the power-law must be compact and central, as seen from the disc.

A further result from the hard state disc variability studies is that on short time-scales, $<1$~s, the direction of disc lags appears to switch so that disc photons start to lag behind power-law variations, with lags of around 1 to 2~ms (see Fig.~\ref{fig:gx339lagvsencomp}).  The natural interpretation of this result is that we are seeing the same X-ray heating of the disc by the power-law as inferred on short time-scales in AGN.  The implied distances between the disc and the power-law emitting region are only a few tens of gravitational radii or even less.  Again, the power-law emitting region appears to be compact, and if it is related to the jet, it must be the base of the jet.

\subsection{Towards a unified picture}
\label{sec:disc:unif}

Discs appear to be inherently unstable in BHXRB hard states, but they are clearly stable in soft states. Does the stabilisation begin within the hard state (perhaps associated with the increase in luminosity prior to the state transition), or does it occur around the state transition?  The harder ($>2$~keV) X-ray variability amplitude also starts to drop continuously during the hard state and then across the state transition, with only a temporary drop when type B QPOs are observed \citep{Belloni2005}.  Since disc variations drive the harder X-ray variability (at least the lower-frequency component, \citealt{Uttley2011}), it seems likely that the stabilisation of disc variability is not strictly linked to the state transition itself but occurs continuously with the overall spectral softening of the source which begins in the hard state, perhaps linked to rises in overall accretion rate that trigger the state transition.   Recent work looking at disc variability in {\it XMM-Newton} data from hard states of SWIFT~J1753.5-0127 strongly supports this picture, showing that the disc is less variable in a softer, higher-luminosity hard state than in a harder, lower-luminosity harder state \citep{Cassatella2012b}.   Data from the {\it Swift} observatory, which enabled monitoring of the soft and hard X-ray variability of SWIFT~J1753.5-0127 throughout a mini-outburst, also supports this picture \citep{Kalamkar2013}.  

The work on disc variability in BHXRBs raises a further question over the very similar patterns seen in AGN variability.  The Seyfert AGN studied with optical/X-ray monitoring mostly show disc-dominated SEDs, i.e. they appear like soft states and not hard states, and yet their discs are clearly variable.  Why do soft-state-like AGN show disc variability while soft state BHXRBs don't?  Perhaps the instability is somehow linked to the radiation pressure instability in discs, since AGN discs should be radiation-pressure dominated while BHXRB discs are more likely to be gas-pressure dominated.  If this is the case, it remains puzzling that soft-state BHXRB discs - which should be {\it more} radiation-pressure dominated - appear to be more stable than hard state discs.

A further question remains over the origin of the higher-frequency variability, seen on sub-second time-scales in BHXRBs and on time-scales of days or less in AGN, where reprocessing of higher-energy photons seems to dominate the disc variability.  The simplest interpretation is that this variability is intrinsic to the corona, hot inner flow, or base of the jet - whichever region produces the power-law emission.  However, this need not be the case: mass accretion variations at higher Fourier frequencies in the disc could be filtered by the more extended disc emissivity profile to suppress the variability of the direct disc emission at these frequencies \citep{Uttley2011,Cassatella2012b}.  It remains an important open question as to whether or not the disc is responsible for driving all the non-jet continuum variability, or only that on longer time-scales.

\section{Synchrotron variability}
\label{sec:sync}

Synchrotron emission from jets has been known as a key emission process in AGN almost since their original discovery in the radio band.  A wealth of observational efforts over the last couple of decades, have shown that jets are almost ubiquitous in accreting stellar mass BHs as well. In BHXRBs, jets are always present throughout the hard state, while their emission is observed to be highly quenched in the soft state \citep{Russell2011}. As we have discussed in the previous Sections, the accretion flow is highly variable, especially in the hard state, where the properties of such variability are also observed to tightly correlate with the jet radio emission \citep{Migliari2005}. Thus, it is perhaps natural to ask if and how such variability is transferred into the jet. The observed jet quenching in the soft state, in which very little variability is observed in the flow, unavoidably broadens the question to the role played by the variability in triggering, powering and influencing the jet itself. 

In this Section, we shall review the observational evidences for jet variability, with an eye on the ongoing theoretical efforts and on the future perspectives.

\subsection{AGN: jet variability, a difficult task}
\label{sec:sync:agn}

The study of jet variability in AGN has been, historically, mostly limited to the highly relativistically-beamed sources such as blazars.  The variability observed in the beamed objects is almost certainly telling us about variability processes produced in the jet, such as shocks which together with the beaming effects can explain the relatively rapid variations seen, especially at very high energies. Since our main focus is on the multiwavelength connections between distinct physical components, we will not cover here the literature on blazar variability which is entirely a jet phenomenon and would require a dedicated Chapter by itself.  However, we will see later how models used to explain blazar variability, in conjunction with accretion-driven variations, can yield useful new ideas for the mechanism that drives the jet emission and variability in BHXRBs.

In non-blazar AGN we have an opportunity to study separately the variations from the jet and other continuum components by studying correlations between variations in other wavebands and the radio emission from the jet.  However, due to the relatively faint nature of non-beamed objects, radio monitoring has been difficult to arrange due to the limited sensitivity of most radio telescopes: this is a particular problem for the weak radio-quiet objects, leading to only a small amount of literature in this area (e.g. \citealt{Mundell2009}).  None-the-less, combined X-ray and radio monitoring campaigns lasting several years were made possible with the observing flexibility of {\it RXTE}.  \citet{Bell2011} reported tentative evidence for an X-ray/radio correlation in the low-luminosity AGN (and possible hard-state analogue) NGC~7213, with 8.4~GHz radio lagging the X-rays by a few weeks (and 4.8~GHz radio lagging by two weeks longer still), assuming that the correlation is real.  Such a lag is consistent with a relatively compact jet in this AGN, which has a black hole mass around $10^{8}$~M$_{\odot}$.  Although NGC~7213 showed significant radio variability and some evidence for an X-ray/radio correlation, the lower mass and higher accretion rate radio-quiet Narrow Line Seyfert~1 AGN, NGC~4051 shows only marginal evidence for radio variability with $\sim 25$\% fractional rms, with no clear evidence for a correlation with X-rays which are highly variable \citep{Jones2011}. However, NGC~4051 does show perhaps the clearest evidence yet for a highly-collimated jet in a radio-quiet AGN, in the form of VLBI images showing a pair of unresolved steep-spectrum radio `hotspots' equidistant from and aligned with the weak, central flat-spectrum radio core \citep{Giroletti2009,Jones2011}.  These findings are important because they strongly suggest that weak jets are present in a type of AGN which is considered to be an analogue of soft state BHXRBs (e.g. see Koerding, this volume).

\subsection{BHXRBs: OIR reveals non-thermal variability}
\label{sec:sync:oir}

The data on radio/X-ray correlations (see K\"{o}rding, this volume) have provided strong evidence for a direct connection between the accretion flow and jets. Past attempts to correlate rapid radio and X-ray variability have confirmed that correlations can be found over different timescales, but that there is little variability in the radio band on timescales shorter than several minutes (e.g. \citealt{Gleissner2004,Nipoti2005,Wilms2007}). This is perhaps not surprising, as the observed radio emission typically comes from at least 10$^{14}$ cm from the black hole, a light travel time of tens of minutes away from the central engine. This is because in a conical jet, the radio emission from the parts of the jet closest to the black hole is absorbed, and the synchrotron self-absorption frequency varies as a function of height along the jet (see Pe'er, this volume). 
Nonetheless, one of the more interesting potential ways to understand the link between the accretion disc and the jet is to probe the variability of the jet and the variability of the disc together, to see how they are coupled. As jets in BHXRBs are known to emit from radio to at least near-infrared (sometimes optical) wavelengths, it is reasonable to look for jet variability at these shorter wavelengths. 

As a figure of merit, it is expected that the spatial scale from which the near-infrared emission should come will be approximately 10$^4$ times smaller than that from which the radio emission comes (i.e. the ratio of wavelengths) giving a light travel time of tens of milliseconds across the IR emitting region. Even with a mildly sub-luminal jet speed, this allows for the likelihood of detecting jet variability on timescales in the 0.1-10 second range, in which most X-ray binaries show most of their variability, without smearing due to light travel-time effects or propagation times along the jet to the region which emits in the IR. 

\begin{figure*}
  \includegraphics[width=0.55\textwidth]{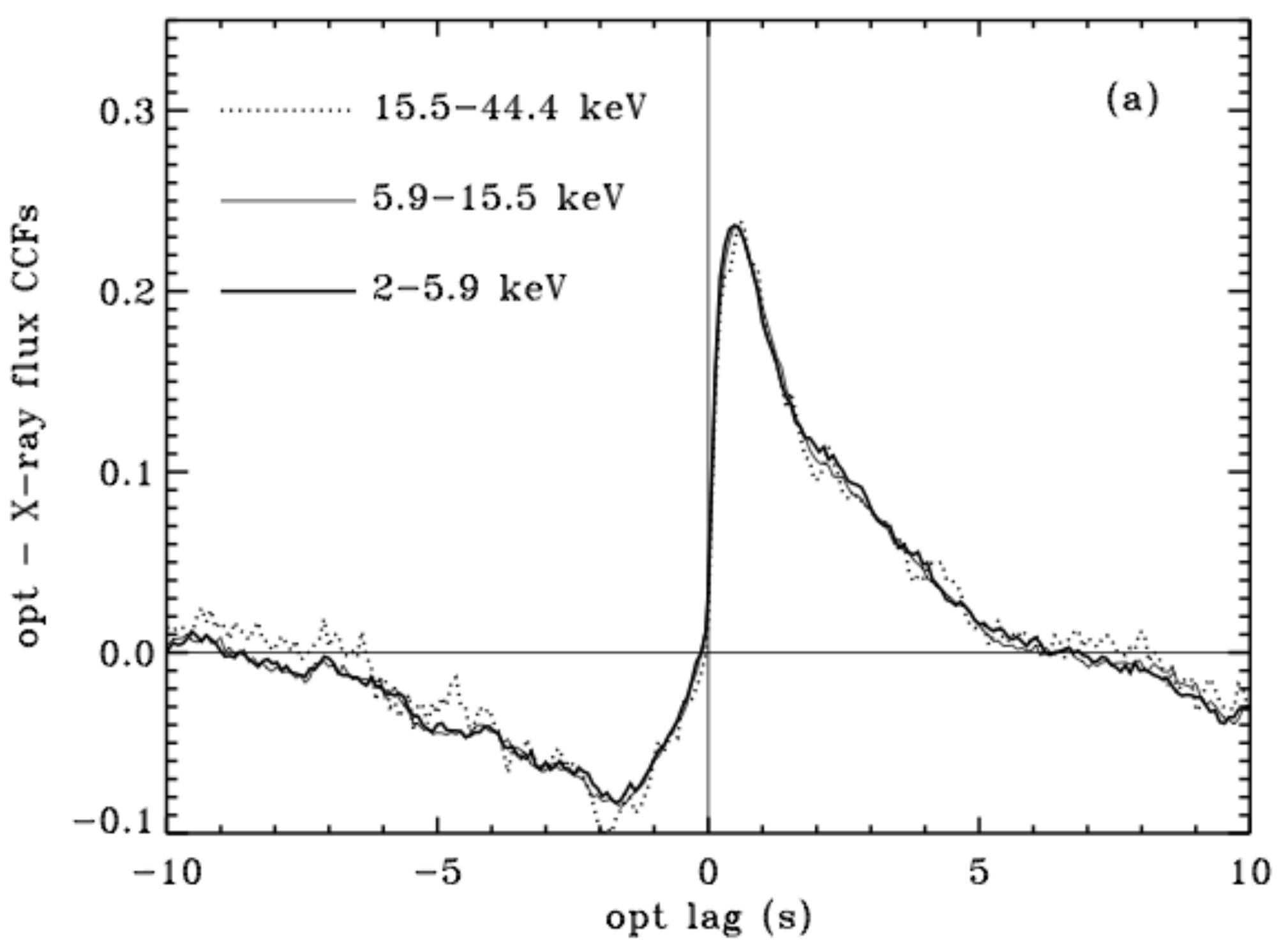}
  \includegraphics[width=0.4\textwidth]{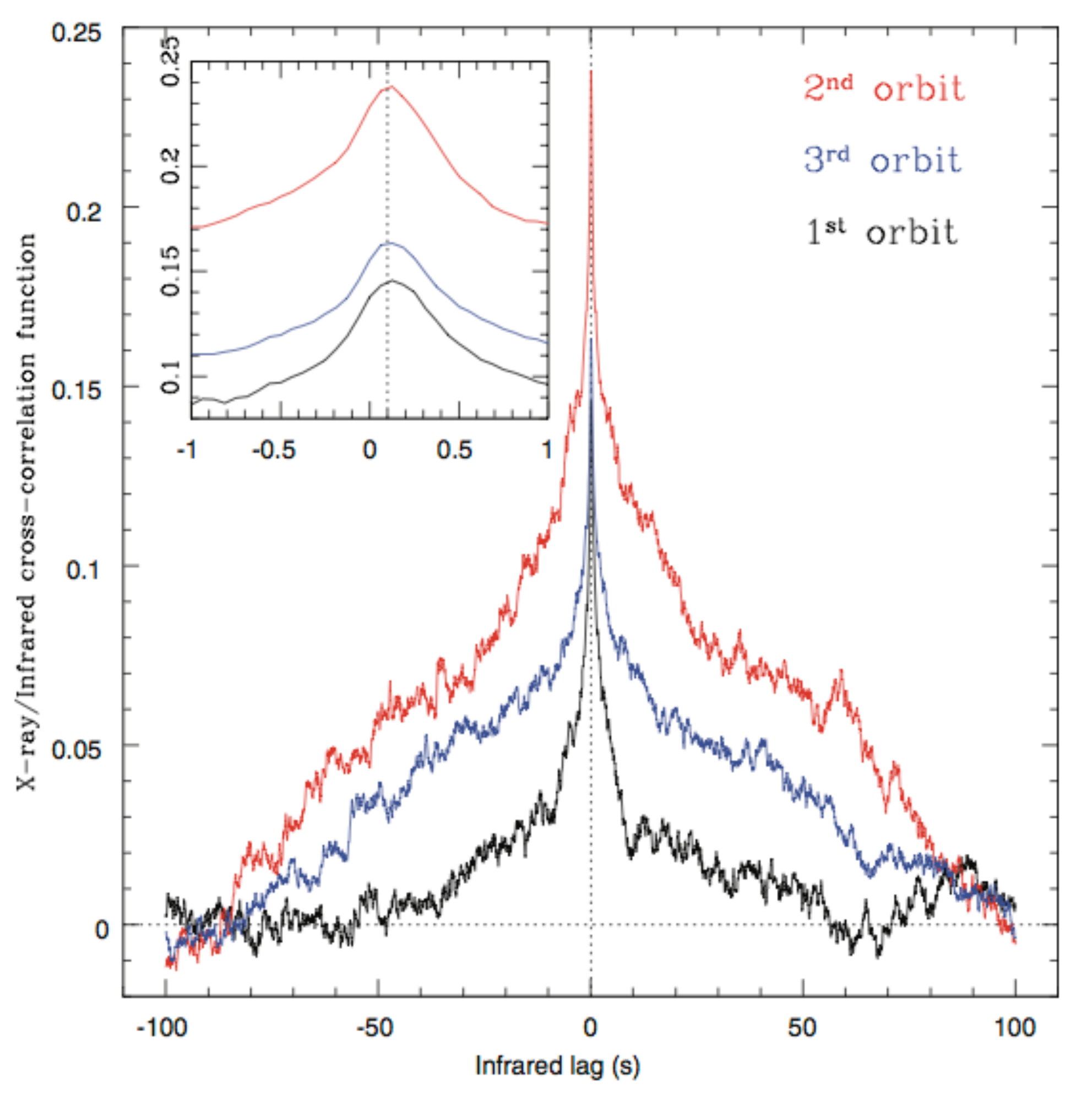}
\caption{Left: Optical/X-ray cross-correlation functions of XTE~J1118+480.  Postive values of the lag correspond to optical lagging X-rays.  Note the complex asymmetric structure, although rapid optical variations which contribute to the sharp peak are clearly lagging the X-rays by a fraction of a second. Figure taken from \citep{Malzac2004}. Right: IR/X-ray cross-correlation functions of GX~339-4.  The shape is much more symmetric than that seen in the optical for XTE~J1118+480, although a significant IR lag of $\sim0.1$~s is still seen, lending strong support to the jet model for the IR variability.}
\label{fig:optirxrayccfs}
\end{figure*}

The first hints for non-thermal variable emission at optical wavelengths came from high-speed simultaneous optical/X-ray photometry of the black-hole candidate XTE~J1118+480.  This source showed a complex, asymmetric cross-correlation function, consisting of a sharp peak corresponding to an optical lag of a few tenths of a second (as it might be expected from reprocessing from the disc), but also an unusual negative value of the cross-correlation function for X-rays lagging behind the optical emission (\citealt{Kanbach2001} and see Fig.~\ref{fig:optirxrayccfs}, left panel). A further complication came with the observations of two other sources with the same technique \citep{Gandhi2008,Durant2008,Gandhi2010}, which also showed very complex optical variability, albeit extremely different from each other. This left several open questions about the actual physical differences between them, and the correct interpretation of the optical/X-ray correlated variability. 

A possible physical explanation for this unusual behavior came from the `common reservoir model' \citep{Malzac2004}, in which part of the optical emission comes from the relativistic jet while the X-ray emission comes from a corona above the accretion disc. Both the jet and the corona stochastically tap the same magnetic energy reservoir, which is also filled stochastically. The fraction of the power that goes into the jet is higher than the fraction that goes into the corona. The correlation is then due to the filling of the reservoir. The time lag of the correlation is assumed to be the time scale on which energy is dissipated in the jet. The anti-correlation is due to the fact that when the jet happens to be tapping more than its usual fraction of the reservoir's energy, the energy is drained sufficiently that the X-ray emission will be suppressed. This model was able to reproduce rather accurately the optical/X-ray CCF of XTE J1118+480, albeit with a large number of free parameters.

An alternative explanation came from \citet{Veledina2011,Veledina2013}, who suggested that the optical emission is produced by synchrotron emission in a magnetised, extended hot accretion flow (for a detailed review of this model, see Poutanen, this volume). A possible caveat to this otherwise promising model is the need for a very large (hundreds of gravitational radii) outer radius of the hot inflow, which is not compatible with what other observations seem to suggest about the location of the X-ray power-law emission and the disc truncation radius \citep[e.g.][]{Uttley2011}. Such large radii however are needed if one wants to obtain significant near-infrared emission from the inflow, whose emission would otherwise be rapidly quenched at wavelengths longer than $\sim$ultraviolet-optical.

Leaving aside the details of each individual model, the scenarios trying to explain the optical variable emission can be divided into two categories, depending on whether such emission is interpreted as coming from the inflow or from the outflow. The two hypotheses differ substantially in the predictions at longer wavelengths: as already mentioned, the inflow is expected to contribute less and less toward longer wavelengths, while the emission from the jet is expected to remain roughly constant - or even to increase (if optically thin) - at wavelengths longer than optical. 

These ambiguities were eventually solved by the first fast IR photometry of a BHXRB \citep{Casella2010}. The IR variability appeared to be strongly correlated with the X-rays, with a very small ($\sim$0.1 seconds) time delay, and a nearly symmetric CCF (see Fig.~\ref{fig:optirxrayccfs}, right panel). Simple calculations and brightness-temperature arguments ruled out a disc-reprocessing origin for this IR variability. Similarly, the inflow origin could be easily ruled out, as in that case an anti-correlation would be expected, let alone the fact that a synchrotron self-Compton, one-zone scenario would require the infrared emission to lead, not to lag the X-rays. Furthermore, and perhaps conclusively, unrealistically large values of the outer hot inflow radius (i.e., of the disc truncation radius) would be needed in order to obtain sufficient infrared emission to explain the data. 

These IR data represented the first unambiguous evidence for sub-second variability in an X-ray binary jet. By making a number of reasonable model-dependent assumptions, the measured time delay between X-rays and infrared photons could be used to estimate either a lower limit of $\Gamma > 2$ for the bulk Lorentz factor of the jet (interpreting the delay in terms of travel time), or a magnetic field intensity of $\sim 10^4$ G at the base of the jet itself (interpreting the delay in terms of cooling time). Furthermore, the Fourier power spectrum of the IR variability showed a clear cutoff at $\sim$1~Hz, which was not visible in the X-ray variability. Interpreting this as a signature of the size of the IR-emitting region in the jet, it was possible to obtain an estimate of $\sim10^{10}$ cm, consistent with that expected for a mildly relativistic jet in these sources.

These quantitative estimates remain as of today largely model dependent, as they will need larger datasets, e.g. by monitoring a single outburst through its accretion rate evolution, in order to suppress the systematic uncertainties caused by the underlying assumptions. Nevertheless, the qualitative yet robust conclusion that the so-called "steady jet" is all but steady remains, challenging the current long-standing view for a steady radio-flat compact jet in the hard state. Perhaps more importantly, the discovered jet variability represents a new promising tool to track matter (and/or internal shocks) through the jet and provides a new tool to measure the geometry and the physics of the jet itself.

\subsection{A variable compact jet}
\label{sec:sync:theory}

A confirmation of the actual complexity of such a variable jet in BHXRBs came from observations at longer wavelengths. \citet{Gandhi2010} reported on four-band WISE simultaneous observations of the BHXRB GX~339-4 over the 3-22 $\mu m$ range, revealing large spectral variability of the jet emission on timescales as short as 11 seconds. The 24-hour long observations showed dramatic variations of the slope and normalization of the spectrum, implying that the spectral break associated with the transition from self-absorbed to optically thin jet synchrotron radiation was varying across the full observed wavelength range. This result suggests that either the magnetic field intensity and/or the size of the acceleration zone above the jet base was being modulated by factors of $\sim$10 on relatively short timescales.

More recently, \citet{Corbel2013} reported on a possibly even more complex jet spectral variation from the same source, as revealed by the Herschel telescope at 70 and 160 $\mu m$ during the outburst decay, with the far-IR flux lying largely above a simple extrapolation of the radio to near-IR spectrum. This result clearly confirmed that the jet spectral emission is far from being as simple and steady as thought in the past (see also Fender \& Gallo and Pe'er, this volume).

A question remains open, as to {\it how} the variability is transferred from the X-ray emitting regions (either a hot inflow or the base of the jet itself) into and along the jet, i.e. in the form of either density or speed fluctuations, or a combination of both.

The answer to this question holds a rather large potential as a tool to investigate jet physics. In fact, internal shocks arising from bulk Lorentz factor fluctuations linked to the variable inflow were used recently as a key ingredient by two jet models. Based on a model by \citet{Spada2001} for blazar variability (itself based on models for Gamma Ray Burst jet emission), \citet{Jamil2010} first tried to reproduce a $\sim$flat radio-to-infrared jet spectrum in BHXRBs by assuming a white-noise distribution of shell bulk Lorentz factor values. The differential kinetic energy between two colliding shells is assumed to be dissipated into the jet, providing the electron re-heating needed to at least partially balance the energy losses (radiative and/or adiabatic; for a detailed discussion of this issue see Pe'er, this volume). More recently, \citet{Malzac2013} generalized this scenario, testing different power-spectral shapes for the variation of the bulk Lorentz factor, namely letting the power-law index of the power spectrum vary from -1 to 2. These works have shown how dissipation in internal shocks driven by flicker noise fluctuations can indeed balance the energy losses, reproducing the observed jet spectra, including the flux amplitude and the location of the self-absorption break frequency. This scenario clearly also predicts strong multiwavelength variability which, assuming some ad-hoc relation between the observed X-ray variability and the speed fluctuations in the jet, can indeed reproduce some of the observed properties (Malzac, private communication).  

These exciting new developments in modelling the jet spectra and variability properties raise further questions about the role of the accretion disc, because as we have discussed in Section~\ref{sec:disc:bhxb}, it appears likely that much of the variation in the X-ray power-law which is correlated with jet variability is actually driven by fluctuations in the blackbody-emitting disc.  Thus, an unstable thin disc may be required to explain the variable flat-spectrum jet emission observed in the hard state.  We can speculate further that the absence of significant compact jet emission in the soft state may be linked to the absence of variability from the accretion disc in the soft state.  That is, without a variable disc there are no variations in bulk Lorentz factor and hence no shocks and consequent reheating to produce emission along the jet.  The jet would retain its bulk kinetic energy however, which may lead to observable signatures in the soft states, but further investigation is needed to test this possibility.

\section{The future}
\label{sec:conc}

A number of multi-wavelength campaigns have shown that variability observed over a broad range of wavelengths and from all physical components is a key ingredient for understanding accretion/ejection physics. Furthermore, advanced spectral-timing techniques are revealing the causal links between the physical components with the X-ray band.  The observational picture is becoming more and more clear, with strong evidence being discovered of how the accretion fluctuations propagate through the accretion disc, possibly into the hot inflow and then into and along the jet.  Future developments will be driven primarily by advances in technology and new instrumentation, together with advances in application of the relevant techniques and modelling.

\subsection{Advances in modelling}
Already a basic phenomenological picture is emerging, with the pieces of the puzzle falling into place.  But the physical models are still substantially behind the data.  Part of the problem is that we are not yet sure what causes the variability, although it appears very likely to be produced in the accretion flow.  However, given the clear phenomenological picture which has emerged in recent years, where we can start to separate out the disc and power-law variations and their associated lags, there is some hope that even simple models for fluctuations propagating through the flow will be useful to fit the data, and that the data will itself put constraints on the models.  There is already a strong indication that variations must originate at relatively small radii to produce the observed lags via viscous propagation, and also that the viscosity parameter $\alpha$ may need to be fairly high to explain the observed lags (\citet{Arevalo2006a,Uttley2011}).  Another recent breakthrough has been the discovery of X-ray reverberation lags (see also Reynolds, this volume).  These lags are more easily modelled since we are dealing with light-travel time effects, and the observed lags and variability time-scales are too short to be significantly effected by the more complex and uncertain effects of propagation in the flow.  Already, the Fe K$\alpha$ reverberation signatures are being modelled using raytraced, general relativistic transfer functions \citep{Cackett2013} and there is justifiable optimism that this approach will allow us to confidently map the inner structure of accreting compact objects for the first time.  The picture that emerges from reverberation studies can naturally fed back into the other models, as it will reveal the geometry of the disc and power-law components which are crucial parts of the picture. 

The approaches to modelling data outlined above are more analytical and `top-down' in terms of the application of the physics.  For example, a source of variability must be assumed to be present in the accretion flow, and possibly injected into the jet, its amplitude and power-spectrum being specified as variables in order to fit the data.  In contrast, a `bottom-up' approach would allow the variability to self-consistently emerge from the physics of the accretion and ejection process.  The current best hope for this emergent picture comes from magneto-hydrodynamic (MHD) simulations of accretion flows and jets, which have made substantial advances in the past decade, due to improvements in the techniques along with significant growth in computing power (e.g. see \citealt{O'Neill2011,McKinney2012} for examples).  These simulations are now beginning to examine the variability that is a natural outcome of, e.g. magnetic instabilities in these flows, and we expect substantial advances in this area over the coming decade, which could allow the first fully-self-consistent modelling of variable accretion and ejection.  As a note of caution however, the coupling of these effects to radiation processes is a challenging task and may present a significant obstacle to modelling the high-accretion rate flows which we observe.  Therefore it is likely that the analytical approach will be the primary driver for understanding the observations for some time to come.

\subsection{Advances in observational capabilities}
Over the next decade we expect significant advances in the quality of multiwavelength data on both long and short time-scales due to a number of important developments.  Firstly, on time-scales of days or longer, (near)-simultaneous high-cadence monitoring of sources will be possible in both the optical and radio, due to the development of synoptic and monitoring survey telescopes in the optical (culminating with the {\it Large Synoptic Survey Telescope} ({\it LSST}|), at the end of the decade), together with {\it Square Kilometre Array} ({\it SKA}) pathfinders (e.g. {\it MeerKAT} and {\it ASKAP}) and ultimately the {\it SKA} itself, which all have significant wide-field monitoring capabilities.  The situation in X-rays is less clear however.  Historically, the development of X-ray All-Sky Monitors has lagged behind that of pointed X-ray telecopes, with the monitoring instruments being seen as secondary `service' instruments to identify new (and relatively bright) X-ray transient sources.  Their sensitivity for monitoring fainter sources, e.g. AGN, is low.  To match the quality of monitoring data that will be provided in the optical and radio in the future, dedicated all-sky or wide-field X-ray monitors need to be developed, perhaps even as separate observatories in their own right.  Possible candidates include the Wide-Field Monitor instrument (WFM) on board the proposed {\it Large Observatory for Timing} ({\it LOFT}), or a sensitive all-sky observatory based on 'Lobster-Eye' optics, but at the time of writing, nothing like this is guaranteed to be flown at the same time as the optical and radio observatories are taking data.  If this remains the case, this would be a very unfortunate missed opportunity: the strong connection between X-ray and optical variability shows that X-ray information is crucial to understanding the observed optical behaviour in accreting compact objects (and almost certainly also the jet behaviour in the radio).

Fortunately, the next decade will see significant advances in observations of rapid variability in all wavebands, which will provide revolutionary new data for the study of accretion and ejection XRBs in particular.  Because of the relatively extended scale of radio emission from XRB jets, we do not expect to see extremely rapid variability in the radio band, although we will have the observational capability to detect it and we should be prepared for surprises.  The most rapid jet variability is expected in the optical/IR and this will be readily detected by a new generation of fast CCD cameras (e.g. HIPERCAM, the successor to ULTRACAM), whose development has been spurred by the study of exoplanetary transits around compact stars, but will naturally lend themselves to fast timing studies of XRBs.  Especially important is the parallel growth in the availability of 4--8~m class telescopes, on which to mount such instruments, since photon count rates are the key to sensitivity to rapid variability.  Many of these new observatories are queue-scheduled (e.g. the {\it South-African Large Telescope}, {\it SALT}) or robotically operated, allowing rapid follow-up of transients.  At the same time, an important advance in optical/IR detector technology is the development of Microwave Kinetic Inductance Detectors (MKIDs) which are able to measure the energies of individual optical/IR photons without using filters or gratings, obtaining similar spectral resolutions to those possible with X-ray CCD instruments (e.g. the ARCONS camera, \citealt{Mazin2013}).  The deployment of MKIDs on large telescopes will thus allow detailed spectral-timing analysis of rapid variability within in the optical-IR range, similar to the developments that have been pioneered in the X-ray band over the past few years.

In the X-ray band, we anticipate continued operation of {\it XMM-Newton}, with its fast EPIC-pn timing capability, into the next decade, allowing the continued application of X-ray spectral-timing techniques to new transients, as well as simultaneous observations with the new fast optical/IR instrumentation.  For simultaneous X-ray/optical/IR spectral-timing, a limiting factor will be the X-ray count rates.  The main limitations of the EPIC-pn instrument when observing bright sources in its fast timing mode are the effects of pileup and deadtime, as well as strong telemetry constraints, meaning the EPIC-pn is effectively limited to observing sources up to $\sim 0.1$~Crab, typically corresponding to BHXRBs in the hard state.  The upcoming {\it ASTROSAT} mission will provide {\it RXTE}-like collecting area and timing capability, together with a larger collecting area that {\it RXTE} at hard energies (which may prove interesting to study any X-ray power-law component of the jet).  Another important development will be the deployment of the {\it Neutron star Interior Composition Explorer} ({\it NICER}) telescope on the International Space Station in 2016.  Although this instrument's core mission is to study X-ray pulsars, its very fast timing-capability coupled with a soft response and CCD-like spectral resolution will allow the detailed spectral-timing studies of hard-state X-ray binaries pioneered by {\it XMM-Newton} to be extended to the more luminous states, where key questions of jet and wind-formation can be studied.
\begin{figure*}
  \includegraphics[width=1.0\textwidth]{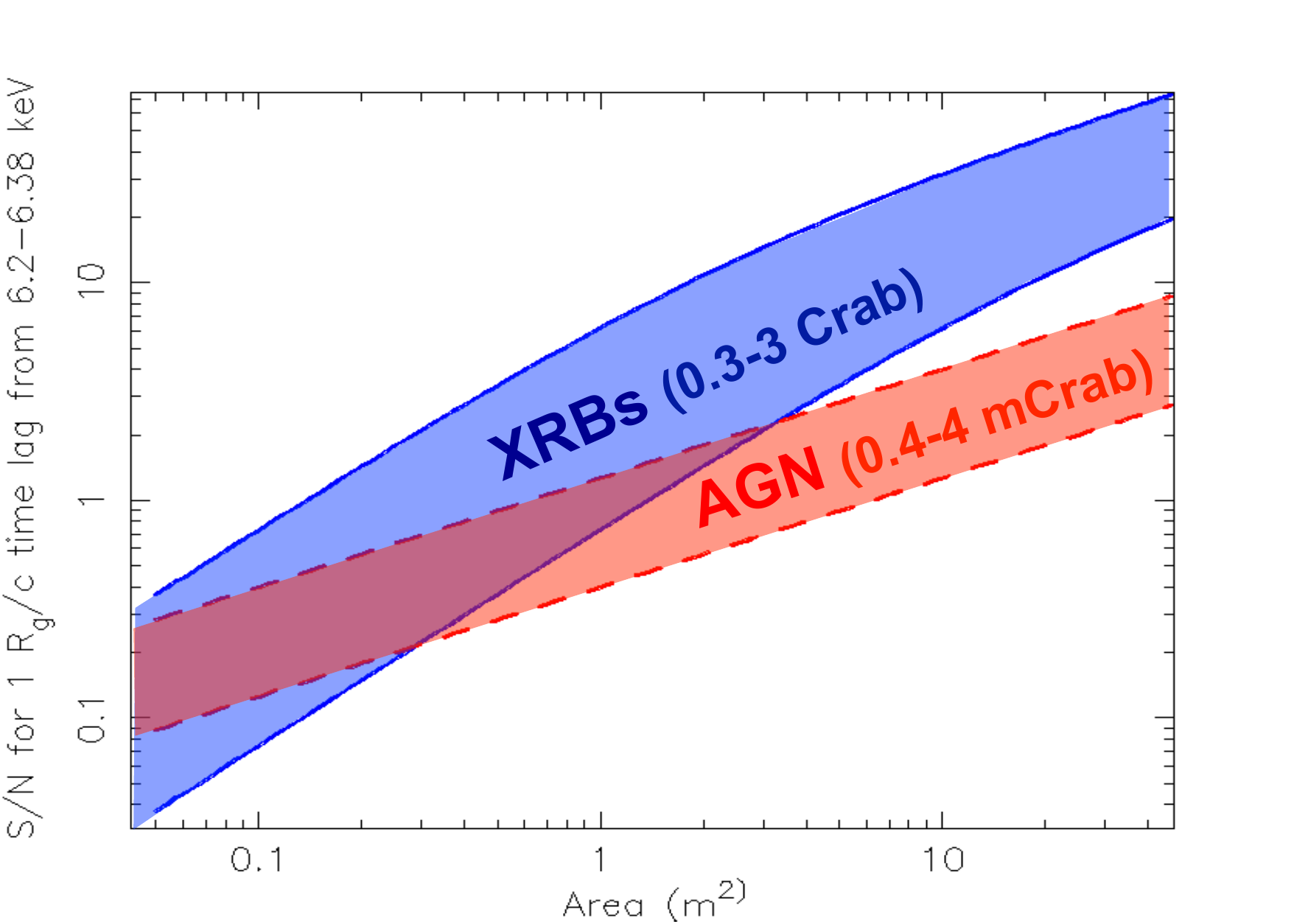}
\caption{The dependence of signal-to-noise on effective area, for the detection of a 1~$R_{\rm g}/c$ Fe~K$\alpha$ lag at high-frequencies (100~Hz in BHXRBs, scaled by black hole mass in AGN).  A typical range of fluxes is assumed for each source class.}
\label{fig:lagareafom}
\end{figure*}

In the longer term, the full potential of X-ray spectral-timing can only be realised at the high count rates facilitated by large-area detectors.  This point is illustrated in Fig.~\ref{fig:lagareafom}, which shows the signal-to-noise ratio expected for the detection of high-frequency time-lags around the Fe~K$\alpha$ line, as a function of detector effective area at that energy.  It is important to note that the sensitivity of lag measurements increases more rapidly for XRBs than for AGN, such that for effective areas above $\sim$1~m$^{2}$, XRB measurements of lags (e.g. for reverberation studies) become significantly better than those for AGN.  The reason for this is that spectral-timing involves the correlation of two light curves, leading to a term in the error which is due to the multiplication of Poisson noise errors from both light curves.  This term dominates the lag-error in the regime where there are few photons per characteristic variability time-scale (i.e. the situation for XRBs), leading to a linear scaling of $S/N$ with count rate (and hence, area) in this regime.  Pushing detector areas to 10~m$^{2}$ - while maintaining CCD-quality spectral resolution - could completely revolutionise the study of accretion/ejection in X-ray binaries by allowing high-resolution reverberation maps of their innermost regions to be made.  The development of large-area silicon-drift detectors puts such detector areas within reach for launch in the 2020s, e.g. the proposed {\it LOFT} mission \citep{Feroci2012}.  Furthermore, in the soft X-ray band the {\it ATHENA} mission, which is now confirmed for launch in the late 2020s, will allow significant advances in studying lags of disc blackbody components and soft X-ray reverberation in AGN \citep{Nandra2013}.

\bibliographystyle{aps-nameyear}      
\bibliography{phil_bibtex}   

%
%



\end{document}